\documentclass[a4paper,11pt]{article}

\usepackage{dsfont}
\usepackage{appendix}
\usepackage{extsizes}
\usepackage{lipsum}
\usepackage[colorlinks]{hyperref}
\usepackage[auth-lg]{authblk}
\usepackage{tcolorbox}
\usepackage{kvoptions}
\usepackage{xparse}
\usepackage{color}
\usepackage{etoolbox}
\usepackage[framemethod=TikZ, xcolor={RGB,table}]{mdframed}
\usepackage{mathtools}
\usepackage{fullpage}
\usepackage{bbold}
\usepackage{enumitem}

\usepackage{fancyhdr}
\usepackage{colonequals}
\usepackage{yfonts}
\usepackage{graphicx}
\usepackage{amstext,amssymb,amsmath,amsthm}
\usepackage[nice]{nicefrac}

\usepackage{mathrsfs}
\usepackage{caption}
\usepackage{centernot}
\usepackage{tikz}
\usetikzlibrary{arrows}
\usepackage{microtype}
\usepackage{algorithm,algpseudocode}
\usepackage[flushmargin,norule]{footmisc}
\usepackage[margin=1.5cm]{geometry}
\usepackage{dsfont}
\usepackage{tcolorbox}

\usetikzlibrary{automata,positioning}

\usepackage{thmtools}
\usepackage{thm-restate}
\usepackage{cleveref}
 \usepackage{optidef}

\tcbuselibrary{theorems}

\hypersetup{linkcolor=blue,citecolor=blue,urlcolor=blue}

\tcbuselibrary{listings,breakable}
\allowdisplaybreaks

\DeclareMathOperator*{\Exp}{\ensuremath{{\mathbb{E}}}}
\DeclareMathOperator*{\Prob}{\ensuremath{\textnormal{Pr}}}
\renewcommand{\Pr}{\Prob}

\newcommand{\prob}[1]{\Pr\paren{#1}}
\newcommand{\card}[1]{|#1|}
\newcommand{\paren}[1]{\left ( #1 \right )}
\newcommand{\bracket}[1]{\left [ #1 \right ]}
\newcommand{\expect}[1]{\Exp\bracket{#1}}

\newcommand{\set}[1]{\ensuremath{\left\{ #1 \right\}}}
\newcommand{\poly}{\mbox{\rm poly}}
\newcommand{\eps}{\varepsilon}

\renewcommand{\algorithmicrequire}{\textbf{Input:}}

\theoremstyle{definition}
\newtheorem{definition2}{Definition}
\theoremstyle{definition}
\newtheorem{example2}[definition2]{Example}
\theoremstyle{definition}

\theoremstyle{definition}
\newtheorem{theorem2}[definition2]{Theorem}
\theoremstyle{definition}
\newtheorem{lemma2}[definition2]{Lemma}
\theoremstyle{definition}
\newtheorem{corollary2}[definition2]{Corollary}
\theoremstyle{definition}
\newtheorem{observation2}[definition2]{Observation}
\theoremstyle{definition}
\newtheorem{claim2}[definition2]{Claim}
\theoremstyle{definition}
\newtheorem{note2}[definition2]{Note}
\theoremstyle{definition}
\newtheorem{remark2}[definition2]{Remark}
\theoremstyle{definition}
\newtheorem{research2}[definition2]{Research Direction}
\theoremstyle{definition}
\newtheorem{conjecture2}[definition2]{Conjecture}
\theoremstyle{definition}

\theoremstyle{definition}

\theoremstyle{definition}
\newtheorem{fact2}[definition2]{Fact}
\theoremstyle{definition}
\newtheorem{result2}[definition2]{Result}

\newenvironment{tbox}{\begin{tcolorbox}[
		enlarge top by=3pt,
		enlarge bottom by=3pt,
		boxsep=0pt,
		left=4pt,
		right=4pt,
		top=10pt,
		arc=0pt,
		boxrule=1pt,toprule=1pt,
		colback=blue!2,
		colframe=blue,
		]
	}
{\end{tcolorbox}}

\newenvironment{tbox2}{\begin{tcolorbox}[
		enlarge top by=3pt,
		enlarge bottom by=3pt,
		breakable,
		boxsep=0pt,
		left=4pt,
		right=4pt,
		top=10pt,
		arc=0pt,
		boxrule=1pt,toprule=1pt,
		colback=blue!2,
		colframe=blue,
		]
	}
	{\end{tcolorbox}}

\newtheorem{mdalg}{Algorithm}

\title{A Note on Rounding Matchings in General Graphs}
\author{Aditi Dudeja\thanks{aditi.dudeja@plus.ac.at. This work is supported by Austrian
Science Fund (FWF): P 32863-N. This project has received funding from the European Research Council (ERC) under the
European Union’s Horizon 2020 research and innovation programme (grant agreement No 947702).}}
\affil{Department of Computer Science, University of Salzburg}
\date{}
\begin{document}
\maketitle
\begin{abstract}
    In this note, we revisit the rounding algorithm of Wajc \cite{Wajc2020}. Wajc gave a fully-adaptive randomized algorithm that rounds a dynamic fractional matching in an \emph{unweighted bipartite graph} to an integral matching of nearly the same value in $O(\poly(\log n,\nicefrac{1}{\varepsilon}))$ update time. We give show that the guarantees of this algorithm hold for \emph{general graphs} as well. Additionally, we show useful properties of this subroutine which have applications in rounding weighted fractional matchings. 
\end{abstract}

In the dynamic matching problem, the graph undergoes edge insertions and deletions, and the algorithm is required to maintain a good approximation to the maximum matching. The goal is to optimize the update time which is the time required to cope with a single edge insertion or deletion. Several dynamic algorithms for approximate maximum matching proceed by computing a fractional matching \cite{BHI15,BHN16,BCH20,BHN17,BlikstadK23} and then rounding it \cite{ArarCCSW18,Wajc2020,BhattacharyaK21,Kiss22,BhattacharyaKSW23}. Fractional matchings have several interesting advantages. They are not only computationally easier to compute, but also certain types of fractional matchings are robust to adversarial edge insertions and deletions \cite{BPT20,AssadiBD22,CST23,JambulapatiJST22}. Thus, there has been a lot of focus on designing efficient rounding algorithms. 

\section{Background on Dynamic Rounding of Fractional Matching}

Arar, Chechik, Cohen, Stein, and Wajc \cite{ArarCCSW18} gave a rounding procedure based on independent sampling: given a dynamic unweighted fractional matching $\vec{x}$, sample every edge with a probability proportional to $x(e)$ and include it in the subgraph $S$. The resulting subgraph $S$ has the property that $\mu(S)\geqslant (1-\varepsilon)\cdot \sum_{e\in E}x(e)$, and is sparse. However, it is only robust to an oblivious adversary. Subsequently, \cite{Wajc2020} gave a dependent version of the above sampling scheme, and showed that the resulting sparsifier was now robust to an adaptive adversary. This rounding procedure, combined with existing fractional matching results, then gave a slew of fully-adaptive dynamic matching algorithms, with varying approximation ratios and update times. However, his analysis only demonstrated that \emph{arbitrary} fractional matchings in bipartite graphs can be rounded to integral matchings of nearly the same value. In another work, Bhattacharya and Kiss \cite{BhattacharyaK21} gave a deterministic rounding procedure for fractional matchings in bipartite graphs. \\ \\
Subsequently, Bhattacharya, Kiss, Sidford, and Wajc \cite{BhattacharyaKSW23} showed that the sampling approach of \cite{ArarCCSW18} can be made robust to output-adaptive adversaries. They also gave deterministic rounding algorithms for bipartite graphs, and for certain types of structured matchings in general graphs. Very recently, Chen, Sidford, and Tu \cite{CST23} showed that the deterministic rounding procedures of \cite{BhattacharyaKSW23} and \cite{ArarCCSW18} also work for general graphs. Curiously, the randomized sparsifier of \cite{ArarCCSW18} also has some additional nice properties, which have applications for rounding weighted matchings. The deterministic sparsifier doesn’t seem to share these additional nice properties. In particular, \cite{ArarCCSW18} show the following theorem for 
dense weighted graphs:

\begin{result2}[Informal \cite{CST23}]\label{result:cst}
There is an output-adaptive randomized algorithm that maintains $(1-\varepsilon)$-approximate maximum weight matching in decremental dense weighted graphs with an update time of $O(\poly(\log n,\frac{1}{\varepsilon}))$.
\end{result2}

Prior to this, all known weighted matching algorithms either had a dependence of $(\nicefrac{1}{\varepsilon})^{O(\nicefrac{1}{\eps})}$, or had a significantly worse approximation ratio than their unweighted counterparts. A crucial step in obtaining \Cref{result:cst}, is the rounding algorithm of \cite{ArarCCSW18} and \cite{BhattacharyaKSW23}. This is also the reason why this algorithm is not fully-adaptive. \\ \\
In their paper, \cite{CST23} left open the question of analysing Wajc’s algorithm, since it is a dependent and consequently, easier to sample from version of \cite{ArarCCSW18}. We show that Wajc’s algorithm also works for general graphs, and additionally show the fully adaptive version of \Cref{result:cst}. In particular, we show the following two results:

\begin{result2}[Informal]
There is a fully-adaptive randomized dynamic rounding algorithm for fractional matchings in general graphs with amortized update time $O(\poly(\log n,\nicefrac{1}{\varepsilon}))$ update time.
\end{result2}

\begin{result2}[Informal]
There is a fully adaptive randomized algorithm that maintains $(1-\varepsilon)$-approximate maximum weight matching in decremental dense weighted graphs with an update time of $O(\poly(\log n,\nicefrac{1}{\varepsilon}))$.
\end{result2}

\section{General Graph Rounding}

We start by defining some notation, then we state Wajc's algorithm. We emphasis that the algorithm remains the same. We merely show that the analysis extends to the case of general (possibly non-bipartite) graphs as well.

\paragraph{Notation} Given a graph $G$, we will use $\mu(G)$ to denote the size of the maximum matching in $G$. Sometimes, we will consider weighted graphs, and in this case, we use $\textsf{mwm}(G)$ to denote the weight of the maximum weight matching. We will use $\vec{x}$ to denote a fractional matching. Given $\vec{x}$, we define 
\begin{align}\label{notation:edges}
    E_{i}&=\set{e\mid x(e)\in \left[(1+\varepsilon)^{-i},(1+\varepsilon)^{-i+1}\right)}\\
    G_i&=(V,E_i)
\end{align}
For a vertex $v$, we define $E_v$ to be the set of edges incident on the vertex $v$. Additionally, given a graph $G$ we have the following notation, which we borrow from \cite{CST23}:
\begin{align*}
    \mathcal{O}&=\set{B\subseteq V\ \big\vert\ \card{B}\text{ is odd}}\\
    \mathcal{O}_{\varepsilon}&=\set{B\subseteq V\ \big\vert\ \card{B}\text{ is odd and }\card{B}\leqslant \nicefrac{1}{\varepsilon} }\\
    \mathcal{P}_{G}&=\set{\vec{x}\in \mathbb{R}^{E}_{\geqslant 0}\ \big\vert\ \sum_{e\in E_v}x(e)\leqslant 1}\\
    \mathcal{M}_{G}&=\set{\vec{x}\in \mathcal{P}_{G}\ \big\vert\ \sum_{e\in G[B]}x(e)\leqslant \frac{\card{B}-1}{2},\forall B\in \mathcal{O}}\\
    \mathcal{M}_{G,\varepsilon}&=\set{\vec{x}\in \mathcal{P}_{G} \ \big\vert\ \sum_{e\in G[B]}x(e)\leqslant \frac{\card{B}-1}{2},\forall B\in \mathcal{O}_{\varepsilon}}
\end{align*}

We will also often use the following fact about matchings.

\begin{fact2}[\cite{AssadiBD22}]\label{obs:smallblossomsuff}
    Suppose $\vec{x}\in\mathcal{M}_{G,\varepsilon}$, then $\frac{\vec{x}}{(1+\varepsilon)}\in \mathcal{M}_{G}$. 
\end{fact2}

\begin{observation2}
    Ignoring the set of edges $E_{\text{low}}=\set{e\in E\mid x(e)\leqslant \frac{\varepsilon^2}{n^2}}$ does not affect $\vec{x}$ by more than a value $\varepsilon$. Consequently, in the notation in (\ref{notation:edges}), we can assume that $i\in \bracket{2\cdot\log \frac{n}{\varepsilon}}$.
\end{observation2}

\begin{algorithm}\caption{\textsc{Sparsification}($G,\vec{x},\varepsilon$)}
\begin{algorithmic}[1]
\For{$i=1,\cdots, \lceil 2\cdot \log \frac{n}{\varepsilon}\rceil $}
\State Edge color the graph $G_i$ using $2\cdot \lceil (1+\varepsilon)^i \rceil$ many colors. \Comment{Coloring Phase}
\State Sample $\min\set{2\cdot \lceil d\cdot (1+\varepsilon)\rceil, 2\cdot \lceil (1+\varepsilon
)^i\rceil}$ colors from $G_i$. Denote sampled edges by $S_i$. \Comment{Sampling Phase}
\EndFor
\State Output $S=\cup_{i} S_i$. 
\end{algorithmic}
\label{alg:sparsification}
\end{algorithm}

The sparsification procedure is stated in \Cref{alg:sparsification}. We first start by giving the following property of this random process, which is a paraphrasing of Lemma 3.2 from \cite{Wajc2020}.

\begin{observation2}\cite{Wajc2020}\label{obs:edgesampleprob}
Suppose $d\geqslant \frac{\log n}{\varepsilon^2}$, for an edge $e$, we have,
\begin{align*}
\frac{\min\set{1,x(e)\cdot d}}{(1+\varepsilon)^2}\leqslant \prob{e\in S}\leqslant \min\set{1, x(e)\cdot d}\cdot (1+\varepsilon).
\end{align*}
Additionally, if $x(e)>\nicefrac{1}{d}$, then $\prob{e\in S}=1$. 
\end{observation2}

\begin{definition2}
    For an edge $e$, we denote $\mathbb{1}_{e}$ to be the indicator random variable for the event $e\in S$. 
\end{definition2}

\begin{definition2}
Observe that the Coloring Phase splits up the graphs $G_i$, and consequently $G$ into monochromatic matchings. Let $M_1,\cdots, M_l$ be these matchings, where $l\leqslant 2\cdot \lceil d\cdot (1+\varepsilon)\rceil\cdot \lceil 2\cdot \log \frac{n}{\varepsilon}\rceil$. For each of these matchings we associate an indicator random variable $\mathbb{1}_{M_i}$ which takes value $1$ if $M_i\in S$ and $0$ otherwise. 
\end{definition2}

\begin{lemma2}(\cite{DP09})\label{lem:samplingworeplace}
Suppose $b_1,b_2,\cdots, b_r$ be any set of $r$ elements. Consider the random process of sampling $k\leqslant r$ objects without replacement from these $r$ elements. Let $X_i$ denote the indicator variable of $b_i$ being included in the sample. Then, $\set{X_i}_{i\in [r]}$ are negatively associated random variables. 
\end{lemma2}

\begin{observation2}\label{obs:negassocv}
Consider a vertex $v$, then the random variables $\set{\mathbb{1}_{e}}_{e\in E_v}$ are negatively associated random variables.
\end{observation2}

We also have the following observation, which 

\begin{observation2}
Consider any odd set $B$, note that the Coloring Phase splits up $G_i[B]$ and consequently, $G[B]$ into monochromatic matchings of size at most $\frac{|B|-1}{2}$. Moreover, all edges on a given matching $M$ are roughly the same weight. Let $M'$ be any such matching in $G_i$ for $i\in \set{1,2,\cdots, \lceil 2\cdot \log \frac{n}{\varepsilon}\rceil}$, and $\mathbb{1}_{M'}$ is an indicator random variable which takes value $1$ if $M'\in S$ in the Sampling Phase. 
\end{observation2}

\begin{lemma2}
Consider any odd set $B$, and let $M'_1,\cdots,M'_j$ be the matchings $G[B]$ has been split up into after the Coloring Phase. Then, $\set{\mathbb{1}_{M'_j}}_{M'_j\in G[B]}$ are negatively associated random variables. 
\end{lemma2}
\begin{proof}
    Consider $G_l$ and recall \Cref{lem:samplingworeplace}. The set of colors, and consequently the matchings correspond to the set of elements we are sampling from in \Cref{lem:samplingworeplace}. Let $M_1,\cdots, M_k$ be the matchings associated with each color (these matchings can be potentially empty, if the colors are not used). In the Coloring Phase, we are sampling a subset of these matchings without replacement. Thus, the random variables, $\set{\mathbb{1}_{M_i}}_{i\in [k]}$ are negatively associated by \Cref{lem:samplingworeplace}. Since subsets of negatively associated random variables are also negatively associated, we have that $\set{\mathbb{1}_{M_i}\mid M_i\neq \emptyset}_{i\in [k]}$ are also negatively associated. Next, the independent union of negatively associated random variables are also negatively associated. Thus, we have, $\mathcal{R}=\set{\mathbb{1}_M\mid M\neq \emptyset}_{M\in G}$ are negatively associated as well. \\ \\
    Finally, consider an odd set $B$, and let $M'_1,\cdots, M'_p$ be the matchings $G[B]$ is split up into. Note that the indicator random associated with these matchings are identical to distinct random variables in $\mathcal{R}$. Consequently, we have, $\set{\mathbb{1}_{M'_i}}_{M'_i\in G[B]}$ are negatively associated as well. 
\end{proof}
Next, we have the following lemma, which is an analog of \Cref{obs:edgesampleprob} for the case of matchings, and in fact directly follows from \Cref{obs:edgesampleprob}.

\begin{observation2}\label{obs:samplingprobmatching}
    Suppose $M\in G_i[B]$, and let $e\in M$, then we have, 
    \begin{align*}
        \frac{d\cdot x(M)}{(1+\varepsilon)\cdot\card{M}}\leqslant \prob{M\in S}\leqslant (1+\varepsilon)^2\cdot \frac{d\cdot x(M)}{\card{M}},
    \end{align*}
if $d<(1+\varepsilon)^{i-1}$, and $\prob{M\in S}=1$ otherwise. 
\end{observation2}

\begin{lemma2}[Chernoff Bounds \cite{DP09}]\label{lem:chernoff}
 Suppose $X_1, \ldots, X_m$ are $m$ negatively associated random variables with range $[0, b]$ each for some $b \geqslant 1$. Let $X:=\sum_{i=1}^m X_i$ and $\mu_L \leqslant \mathbb{E}[X] \leqslant \mu_H$. Then, for any $\delta>0$,
$$
\operatorname{Pr}\left(X>(1+\delta) \cdot \mu_H\right) \leqslant \exp \left(-\frac{\delta^2 \cdot \mu_H}{(3+\delta) \cdot b}\right) \quad \text { and } \quad \operatorname{Pr}\left(X<(1-\delta) \cdot \mu_L\right) \leqslant \exp \left(-\frac{\delta^2 \cdot \mu_L}{(2+\delta) \cdot b}\right)
$$
\end{lemma2}

\begin{lemma2}\label{lem:propofsparsification}
Suppose $\vec{x}\in \mathcal{M}_{G}$ is input to \textsc{Sparsification}(), and $d\geqslant \frac{\log n}{\varepsilon^4}$, then, with high probability, there is a vector $\vec{y}\subseteq \text{supp}(S)$ such that:
\begin{enumerate}[label=\emph{\alph*})]
\item For all vertices $v\in V$, $(1-\varepsilon)\cdot \sum_{e\in E_v}x(e)-\varepsilon\leqslant \sum_{e\in E_v}y(e)\leqslant (1-\varepsilon)\cdot\sum_{e\in E_v} x(e)+\varepsilon$. \label{prop:b}
\item \label{prop:c} For all odd sets $B$ with $\card{B}\leqslant \nicefrac{1}{\varepsilon}$, 
\begin{align*}
(1-\varepsilon)\cdot \sum_{e\in G[B]}x(e)-\frac{\varepsilon\cdot (\card{B}-1)}{2}\leqslant \sum_{e\in G[B]}y(e)\leqslant (1-\varepsilon)\cdot \sum_{e\in G[B]}x(e)+\frac{\varepsilon\cdot \card{B}}{2}.
\end{align*}
\end{enumerate}
Moreover, if $\sum_{e\in E}x(e)\geqslant (1-\varepsilon)\cdot \mu(G)$, then with a constant probability, we have, that $\vec{y}$ satisfies:
\begin{align}
  \sum_{e\in E}y(e)\geqslant (1-\varepsilon)\cdot \sum_{e\in E}x(e).\label{prop:a}
\end{align}
\end{lemma2}
The above lemma essentially states that with a constant probability there is a fractional matching $\vec{y}\subseteq \text{supp}(S)$ with $\sum_{e\in E} y(e)\geqslant (1-\varepsilon)\cdot \sum_{e\in E} x(e)$ satisfying small odd set constraints.

\begin{proof}
We first define a flow $\vec{y}$ on the support of $S$. Then, we show \ref{prop:b} and \ref{prop:c} are satisfied for $\vec{y}$ with high probability. Finally, we conditioned on \ref{prop:b} and \ref{prop:c}, we show that $\sum_{e\in E}y(e)\geqslant (1-\varepsilon)\cdot \sum_{e\in E}x(e)$ with a constant probability. Let $S_v=\set{e\ni v\mid x(e)<\frac{1}{d}}$. We now define the flow as follows, for all $v\in V$, $e\in  S_v$,
\begin{align*}
    y(e)=\frac{1}{d}\cdot \mathbb{1}_{e}\cdot (1-4\varepsilon)
\end{align*}
On the other hand, for $e\notin S_v$, we let $y(e)=x(e)$. We first show \ref{prop:b}. Consider an edge $e\in S_v$, from \Cref{obs:edgesampleprob}, we have,
\begin{align*}
   x(e)\cdot (1-3\varepsilon)\leqslant \expect{y(e)}\leqslant x(e)\cdot(1-2\varepsilon)
\end{align*}
Consequently, for a vertex $v$, we have,
\begin{align*}
    (1-3\varepsilon)\cdot \sum_{e\in S_v} x(e)\leqslant \expect{\sum_{e\in S_v}y(e)}\leqslant (1-2\varepsilon)\cdot \sum_{e\in S_v}x(e)
\end{align*}
Note that to show concentration, it is sufficient to consider the set of edges $S_v$. This is because of \Cref{obs:edgesampleprob}. Next, observe that for $e\in S_v$, $y(e)\leqslant \frac{1}{d}$. Finally, by \Cref{obs:negassocv}, we have, that $\set{y(e)}_{e\in S_v}$ are negatively associated. Thus, by Chernoff bound (\Cref{lem:chernoff}), we have, the following upper tail bound by picking $\mu_H=(1-2\varepsilon)\cdot \sum_{e\in S_v}x(e)$, and $\delta = \frac{\varepsilon}{(1-2\varepsilon)\cdot \sum_{e\in S_v}x(e)}$,
\begin{align*}
    \prob{\sum_{e\in S_v}y(e)\geqslant (1-2\varepsilon)\cdot \sum_{e\in S_v}x(e)+\varepsilon}&\leqslant \exp\paren{-\frac{\paren{\frac{\varepsilon}{(1-2\varepsilon)\cdot \sum_{e\in S_v}x(e)}}^2\cdot \paren{(1-2\varepsilon)\cdot \sum_{e\in S_v}x(e)}}{\paren{3+\frac{\varepsilon}{(1-2\varepsilon)\cdot\sum_{e\in S_v}x(e)}}\cdot \frac{\varepsilon^4}{\log n}}}\\
    &= O\paren{\frac{1}{n^{\nicefrac{1}{\varepsilon}}}}
\end{align*}
Similarly, substituting $\mu_L=(1-3\varepsilon)\cdot \sum_{e\in S_v}x(e)$ and $\delta=\frac{\varepsilon}{(1-3\varepsilon)\cdot \sum_{e\in S_v}x(e)}$,
\begin{align*}
    \prob{\sum_{e\in S_v}y(e)\leqslant (1-3\varepsilon)\cdot \sum_{e\in S_v}x(e)-\varepsilon}&\leqslant \exp\paren{-\frac{\paren{\frac{\varepsilon}{(1-3\varepsilon)\cdot \sum_{e\in S_v}x(e)}}^2\cdot \paren{(1-3\varepsilon)\cdot \sum_{e\in S_v}x(e)}}{\paren{2+\frac{\varepsilon}{(1-3\varepsilon)\cdot\sum_{e\in S_v}x(e)}}\cdot \frac{\varepsilon^4}{\log n}}}\\
    &= O\paren{\frac{1}{n^{\nicefrac{1}{\varepsilon}}}}
\end{align*}
Finally, since $\sum_{e\in E_v\setminus S_v}y(e)=(1-4\eps)\cdot \sum_{e\in E_v\setminus S_v}x(e)$ for all $v\in V$, we can conclude that with high probability,
\begin{align*}
(1-4\varepsilon)\cdot \sum_{e\in E_v}x(e)-\varepsilon\leqslant \sum_{e\in E_v}y(e)\leqslant (1-3\varepsilon)\cdot\sum_{e\in E_v}x(e)-\varepsilon
\end{align*}
We now show \ref{prop:c}. Consider an odd set $B$, and let $M'_1,\cdots, M'_k$ denote the set of matchings $G[B]$ is split up into. As before, we can consider only matchings in $G_i$, where $d<(1+\varepsilon)^{i-1}$ due to \Cref{obs:samplingprobmatching}.
\begin{align*}
    \expect{\sum_{e\in G[B]}y(e)}=\expect{\sum_{i=1}^k y(M'_i)}\leqslant \sum_{i=1}^k x(M'_i)\cdot (1-4\varepsilon)\leqslant \sum_{e\in G[B]}x(e)\cdot (1-4\varepsilon)
\end{align*}
Similarly, we have,
\begin{align*}
    \expect{\sum_{e\in G[B]}y(e)}=\expect{\sum_{i=1}^k y(M'_i)}\geqslant \sum_{e\in G[B]}x(e)\cdot (1-5\varepsilon)
\end{align*}
Additionally, observe that $y(M'_i)$ can take value at most $\frac{\card{M'_i}}{d}\leqslant \frac{\card{B}-1}{2\cdot d}$. Thus, using $\mu_H=\sum_{e\in G[B]}x(e)\cdot (1-4\varepsilon)$, $\delta= \frac{\varepsilon\cdot(\card{B}-1)}{\sum_{e\in G[B]}x(e)\cdot (1-4\varepsilon)}$, we have,
\begin{align*}
    \prob{\sum_{e\in G[B]}y(e)\geqslant \sum_{e\in G[B]}x(e)\cdot (1-4\varepsilon)+\frac{\varepsilon\cdot (\card{B}-1)}{2}}&\leqslant \exp\paren{-\frac{\paren{\frac{-\varepsilon^2\cdot(\card{B}-1)^2}{4\cdot\paren{\sum_{e\in G[B]}x(e)\cdot (1-4\varepsilon)}^2}}\cdot (1-4\varepsilon)\cdot \sum\limits_{e\in G[B]}x(e)}{\paren{3+\frac{\varepsilon}{\sum_{e\in G[B]}x(e)\cdot (1-3\varepsilon)}}\cdot \frac{\card{B}-1}{2\cdot d}}}\\
    &= O\paren{\frac{1}{n^{\nicefrac{1}{\varepsilon^2}}}}.
\end{align*}
Using the fact that $\sum_{e\in G[B]}x(e)\leqslant \frac{\card{B}-1}{2}$ and $d\geqslant \nicefrac{\log n}{\varepsilon^4}$, we have the second equality. Finally, we union bounding over all odd sets of size at most $\frac{1}{\varepsilon}$, we have the upper bound. For the lower bound, consider $\mu_L=(1-5\varepsilon)\cdot \sum_{e\in G[B]}x(e)$, and $\delta=\frac{\varepsilon\cdot (\card{B}-1)}{(1-4\varepsilon)\sum_{e\in G[B]}x(e)}$, 
\begin{align*}
    \prob{\sum_{e\in G[B]}y(e)\geqslant \sum_{e\in G[B]}x(e)\cdot (1-4\varepsilon)-\frac{\varepsilon\cdot (\card{B}-1)}{2}}&\leqslant \exp\paren{-\frac{\paren{\frac{-\varepsilon^2\cdot(\card{B}-1)^2}{4\cdot\paren{\sum_{e\in G[B]}x(e)\cdot (1-4\varepsilon)}^2}}\cdot (1-4\varepsilon)\cdot \sum\limits_{e\in G[B]}x(e)}{\paren{3+\frac{\varepsilon}{\sum_{e\in G[B]}x(e)\cdot (1-3\varepsilon)}}\cdot \frac{\card{B}-1}{2\cdot d}}}\\
    &\leqslant O\paren{\frac{1}{n^{\nicefrac{1}{\varepsilon}^2}}}.
\end{align*}
Taking a union bound over all odd sets of size at most $\frac{1}{\varepsilon}$, we have \ref{prop:c}. We now move on to proving \Cref{prop:a}. First, let $\mathcal{E}_{b,c}$ denote the event that \ref{prop:b} and \ref{prop:c} occur. We have,
\begin{align*}
    \prob{\mathcal{E}_{b,c}}\geqslant 1-O\paren{\frac{1}{n^{\nicefrac{1}{\eps^2}}}}.
\end{align*}
Additionally, we have,
\begin{align*}
    \expect{\sum_{e\in E}y(e)}\geqslant (1-4\varepsilon)\cdot \sum_{e\in E} x(e).
\end{align*}
By law of total probability, we have,
\begin{align*}
    \expect{\sum_{e\in E}y(e)\Bigm\vert \mathcal{E}_{b,c}}\cdot \prob{\mathcal{E}_{b,c}}+ \expect{\sum_{e\in E}y(e)\Bigm\vert \lnot\mathcal{E}_{b,c}}\cdot \prob{\lnot\mathcal{E}_{b,c}}&\geqslant (1-4\varepsilon)\cdot \sum_{e\in E} x(e)\\
    \expect{\sum_{e\in E}y(e)\Bigm\vert \mathcal{E}_{b,c}}+n^2\cdot O\paren{\frac{1}{n^{\nicefrac{1}{\varepsilon}}}}&\geqslant (1-4\varepsilon)\cdot \sum_{e\in E} x(e)
\end{align*}
Thus, we have, $\expect{\sum_{e\in E}y(e)\Bigm\vert \mathcal{E}_{b,c}}\geqslant (1-5\varepsilon)\cdot \sum_{e\in E}x(e)$. Let $\vec{x}^*\in \mathcal{M}_{G}$ be any optimum fractional matching. Conditioned on $\mathcal{E}_{b,c}$, we have $\vec{y}\in \mathcal{M}_{G,\varepsilon}$. Thus, from \Cref{obs:smallblossomsuff}, we have, $\sum_{e\in E}y(e)\leqslant (1+\varepsilon)\cdot \sum_{e\in E}x^*(e)$. Therefore, the random variable $(1+\varepsilon)\cdot \sum_{e\in E} x^*(e)-y(e)$ is non-negative, and moreover, by our assumption that $\sum_{e\in E}x(e)\geqslant (1-\varepsilon)\cdot\sum_{e\in E}x^*(e)$, we have,
\begin{align*}
\expect{(1+\varepsilon)\cdot\sum_{e\in E} x^*(e)-y(e)\Bigm\vert \mathcal{E}_{b,c}}\leqslant 4\varepsilon\cdot \sum_{e\in E}x^*(e)
\end{align*}
Thus, by Markov inequality, we have, 
\begin{align*}
    \prob{(1+\varepsilon)\cdot\sum_{e\in E} x^*(e)-y(e)\geqslant 8\varepsilon\cdot \sum_{e\in E}x^*(e)\Bigm\vert\mathcal{E}_{b,c}}\leqslant \frac{1}{2}
\end{align*}
This implies,
\begin{align*}
    \prob{\sum_{e\in E}y(e)\leqslant (1-8\varepsilon)\cdot\sum_{e\in E}x^*(e)\Bigm\vert \mathcal{E}_{b,c}}\leqslant \frac{1}{2}
\end{align*}
Thus, the total probability of either \ref{prop:b},\ref{prop:c}, or \Cref{prop:a} not happening is upper bounded as follows:
\begin{align*}
    \prob{\lnot \mathcal{E}_{b,c}}+\prob{\sum_{e\in E}y(e)\leqslant (1-8\varepsilon)\cdot \sum_{e\in E}x^*(e)\ \Bigm\vert\  \mathcal{E}_{b,c}}\leqslant \frac{1}{2}+O\paren{\frac{1}{n^{\nicefrac{1}{\varepsilon}^2}}}.
\end{align*}
This proves our lemma. 
\end{proof} 

\begin{corollary2}
    Let $\vec{x}\in \mathcal{M}_{G}$ be such that $\sum_{e\in E}x(e)\geqslant(1-\varepsilon)\cdot \mu(G)$. Let $S$ be the output of \textsc{Sparsification}($G,\vec{x},\varepsilon$). Then, $\mu(S)\geqslant (1-\varepsilon)\cdot \sum_{e\in E}x(e)$. 
\end{corollary2}
\begin{proof}
    Note that by \Cref{prop:a}, we know that there is $\vec{y}\subseteq \text{supp}(S)$ such that $\sum_{e\in E}y(e)\geqslant (1-\varepsilon)\cdot \sum_{e\in e}x(e)$. Additionally, from \Cref{lem:propofsparsification}\ref{prop:b} we know that for every vertex $v$,
    \begin{align*}
        \sum_{e\in E_v} y(e)\leqslant (1-\varepsilon)\cdot\sum_{e\in E_v} x(e)+\varepsilon\leqslant 1
    \end{align*}
    Finally, we know that for any odd set $B$ with $\card{B}\leqslant \nicefrac{1}{\varepsilon}$,
    \begin{align*}
        \sum_{e\in G[B]}y(e)\leqslant (1-2\varepsilon)\cdot \sum_{e\in G[B]} x(e)+\frac{\varepsilon\cdot\card{B}}{2}\leqslant (1-2\varepsilon)\cdot \frac{\card{B}-1}{2}+\frac{\varepsilon\cdot(\card{B}-1)}{2}\leqslant (1-\varepsilon)\cdot\frac{\card{B}-1}{2}
    \end{align*}
Thus, we have, $\frac{\vec{y}}{1+\varepsilon}\in \mathcal{M}_{G}$. 
\end{proof}

Next, we show two properties of the \Cref{alg:sparsification} that \cite{Wajc2020} gives in his paper.

\begin{observation2}\cite{Wajc2020}\label{obs:sizeofsparsifier}
Let $\vec{x}\in \mathcal{M}_{G}$ be a fractional matching of $G$, and let $S$ be the output of \textsc{Sparsification}($G,\vec{x},\varepsilon$). Then, $\card{E(H)}= O(d\cdot\mu(G)\cdot \log\frac{n}{\varepsilon})$.
\end{observation2}
\begin{observation2}\cite{Wajc2020}\label{obs:sparsifierupdatetime}
    Given a dynamic fractional matching $\vec{x}$, \Cref{alg:sparsification} maintains a sparsifier $S$ of $\vec{x}$ satisfying \Cref{lem:propofsparsification} with update time $O(\poly(\log n,\nicefrac{1}{\varepsilon}))$. Moreover, the guarantees of \Cref{alg:sparsification} hold against an adaptive adversary. 
\end{observation2}

\begin{lemma2}\label{lem:duanpettie}\cite{DP14}
    There is a static algorithm, \textsc{Static-Match}() that computes a $(1-\varepsilon)$-approximation to the maximum weight matching in $O(m\cdot \nicefrac{1}{\varepsilon}\cdot \log\nicefrac{1}{\varepsilon})$ time. 
\end{lemma2}

\begin{theorem2}
There is a randomized dynamic rounding algorithm for general graphs with amortized update time $O(\poly(\log n,\nicefrac{1}{\varepsilon}))$. The algorithm is robust to adaptive adversaries.
\end{theorem2}
\begin{proof}
    Suppose $\vec{x}$ is the initial fractional matching with $\sum_{e\in E}x(e)\geqslant (1-\varepsilon)\cdot \textsf{mwm}(G)$, then \Cref{alg:sparsification} initializes the sparsifier in $O(m\cdot \poly(\log n, \nicefrac{1}{\varepsilon})$ total time, and outputs the graph $S$. From \Cref{obs:sizeofsparsifier}, we know that $\card{E(S)}=O(d\cdot \mu(G)\cdot \log \frac{n}{\varepsilon})$. Thus, we can run a static matching algorithm (say, \cite{DP14}) to get a matching $M$ with $\card{M}\geqslant (1-\varepsilon)\cdot \mu(S)$. By \Cref{lem:propofsparsification} and \Cref{obs:smallblossomsuff}, we have,
    \begin{align*}
        \mu(S)\geqslant (1-\varepsilon)^2\sum_{e\in E}x(e)\geqslant (1-\varepsilon)^2\cdot \mu(G)
    \end{align*}
Now, we have to argue how the rounding algorithm deals with updates to the input, that is, the fractional matching. Suppose the input fractional matching $\vec{x}$ is modified. Then, note that by \Cref{obs:sparsifierupdatetime}, in $O(\poly(\log n, \nicefrac{1}{\varepsilon}))$ time, we can update the sparsifier $S$ to reflect the change in the fractional matching. As for the integral matching $M$, if the deleted edge $e$ is in $M$, then we remove it from $M$. If at any point, $\card{M}\leqslant (1-2\cdot \varepsilon)\cdot \sum_{e\in E}x(e)$, then we recompute $M$ by running \cite{DP14} on $S$. The amortized time now, will be $O(\poly(\log n,\nicefrac{1}{\varepsilon}))$, since we waited for at least $\varepsilon\cdot \mu(G)$ updates before recomputing the matching. The runtime of \cite{DP14} is $O(d\cdot \mu(G)\cdot \poly(\log n,\nicefrac{1}{\varepsilon}))$ by \Cref{obs:sizeofsparsifier}. Finally, \Cref{alg:sparsification} is robust to adaptive adversaries, and \cite{DP14} is a deterministic algorithm. Consequently, the rounding algorithm is robust to adaptive adversaries. 
\end{proof}

\section{Application}

In this section, our goal will be to show the following theorem.

\begin{restatable}{theorem2}{densegraphs}\label{thm:main}
There is a randomized fully adaptive algorithm maintaining a $(1-\varepsilon)$-approximate integral weighted matching in a decremental graph with amortized update time $\Tilde{O}(\varepsilon^{-41}+\frac{n^2}{m}\cdot \varepsilon^{-6})$.
\end{restatable}
In their work, \cite{CST23} proved the same theorem, but with guarantees against an output-adaptive adversary. In this section, we will strengthen this to a fully adaptive adversary. Our main tool will be as in the case of \cite{CST23}, to use the \emph{entropy regularized matching}. Their proof strategy was as follows:
\begin{enumerate}
    \item They first consider the sparsification algorithm $\mathcal{A}$ of \cite{ArarCCSW18}, which is used to round \textbf{unweighted fractional matchings}.  They showed that $\mathcal{A}$ output a sparsifier $S$ which was also a degree sparsifier. A degree sparsifier is a subgraph, which satisfies a stronger version of \Cref{weakdegsparsifier}. This algorithm is robust against output-adaptive adversaries.
    \item Next, they considered the \emph{modified entropy regularized matching} $\vec{x}$ (see \Cref{def:modifiedentropy}) and showed the following:
    \begin{enumerate}
        \item We have, $\sum_{e\in E} x(e)\cdot w(e)\geqslant (1-\varepsilon)\cdot \textsf{mwm}(G)$.
        \item When $\vec{x}$ is fed into $\mathcal{A}$, then the sparsifier $S$ output by the algorithm not only preserves most of its weight. 
        \item The fractional matching $\vec{x}$ can be computed in $O(m\cdot \poly(\log n,\nicefrac{1}{\varepsilon}))$ time. 
    \end{enumerate}
They combine all of these facts to show the output adaptive version of \Cref{thm:main}. 
\end{enumerate}
We follow a similar strategy: we show the sparsifier output by \Cref{alg:sparsification} is a weak-degree sparsifier. We consider \emph{entropy regularized matching} problem, and show that when $\vec{x}$ is fed into \textsc{Sparsification}(), then most of its weight is preserved. Now, since \textsc{Sparsification}() is fully-adaptive, we are able to prove a stronger version of the corresponding theorem of \cite{CST23}. In the subsequent section, we will start by giving the definition of a weak degree sparsifier and show \textsc{Sparsification}() outputs a graph satisfying these properties.

\subsection{Weak $\varepsilon$-degree Sparsifier}

We consider the modified definition of a degree sparsifier from \cite{CST23}.
\begin{definition2}\label{weakdegsparsifier}[Weak $\eps$-degree sparsifier]
Suppose $\vec{x}$ is a fractional matching of $G$, then $H$ is an $s$-sparse, weak $\varepsilon$-degree sparsifier of $G$ if $\card{H}\leqslant s\cdot \sum_{e\in E}x(e)$, and moreover, there exists a fractional matching $x^{(H)}\in \mathcal{M}_G$ supported on $H$, such that:
\begin{enumerate}
    \item For all $v\in V$, we have, $x^{H}(v)\geqslant x(v)-\varepsilon$,
    \item For all odd sets $B\in \mathcal{O}^{\varepsilon}_{G}$, we have, $\sum_{e\in G[B]}x^{H}(e)\geqslant \sum_{e\in G[B]}x(e)-\varepsilon\cdot \frac{\card{B}-1}{2}$, and
    \item If $\sum_{e\in E}x(e)\geqslant (1-\varepsilon)\cdot \mu(G)$, then $\sum_{e\in E}x^{H}(e)\geqslant (1-\varepsilon)\cdot \sum_{e\in E}x(e)$,
\end{enumerate}
\end{definition2}
\begin{corollary2}\label{cor:sparsificationandweakdegreesparse}
    From \Cref{lem:propofsparsification}, we can deduce that if $\vec{x}$ is input to \textsc{Sparsification}() with parameter $\varepsilon$ and $d\geqslant \frac{\log n}{\varepsilon^4}$, then the output $S$ is a $O(\log n)$-sparse, weak $\varepsilon$-degree sparsifier of $G$ with probability at least $0.5$. 
\end{corollary2}

In the subsequent section we define the fractional matching problem we will be solving.

\subsection{Entropy Regularized Matching}

The next part is, to show the properties of \textbf{entropy regularized matching problem}. Note that \cite{CST23} considered the \textbf{modified entropy regularized matching problem}, given in \Cref{def:modifiedentropy}. We consider a different version due to the fact that our sparsifier is now weaker, so to prove the required guarantees, we need to solve a slightly different version of the problem than \cite{CST23}.

\begin{definition2}\label{def:entropyregmatching}
The \emph{entropy regularized matching} problem is the problem of finding a fractional matching $\vec{x}\in \mathcal{M}_{G,\varepsilon}$, maximizing the following objective function:
\begin{align*}
g(\vec{x})=\sum_{e\in E} x(e)\cdot w(e)+\delta\cdot\sum_{e\in E}w(e)\cdot x(e)\cdot\log\frac{\gamma}{w(e)\cdot x(e)}
\end{align*}
\end{definition2}

\begin{remark2}
    Note that \cite{CST23} solve the same objective function as in \Cref{def:entropyregmatching}, but over the polytope $\mathcal{M}_G$. 
\end{remark2}

The next observation is a modified version of Lemma 4.8 in \cite{CST23}.

\begin{observation2}\label{obs:entropyregmwm}
    If $\delta\leqslant \nicefrac{\varepsilon}{8\cdot\log nW}$ and $\textsf{mwm}(G)\leqslant \gamma\leqslant \textsf{mwm}(G)\cdot n\cdot W$, then and suppose $\vec{x}^*\in \mathcal{M}_{G,\varepsilon}$ is the optimal solution to $g$, then $g(\vec{x^*})\leqslant (1+\varepsilon)\cdot \textsf{mwm}(G)$.
\end{observation2}

Let $Z^{\delta}_{E,\gamma}$ denote the optimal value of the entropy regularized matching over $\mathcal{M}_{G,\varepsilon}$. Moreover, let $\vec{x}^*\in \mathcal{M}_{G,\varepsilon}$ be the unique matching realizing this value. Let $w^*=\sum_{e\in E}w(e)\cdot x^*(e)$. We have the following lemma, which is an adaptation of Lemma 6.23 of \cite{CST23}.
\begin{lemma2}\label{optimalprimal}
For an edge $e=(u,v)$ let $\mathcal{O}_{\varepsilon}^e=\set{B\in \mathcal{O}_{\varepsilon}\bigm\vert e\in G[B]}$. There exist a pair of vertex duals $\vec{y}$ and odd set duals $\vec{z}$ such that for every edge $e$, we have,
\begin{align*}
x^*(e)=2^{\frac{1}{\delta}-1-\frac{s_e}{\delta\cdot w(e)}+\log\frac{\gamma}{w(e)}},
\end{align*}
where $s_e=y_u+y_v+\sum_{B\in \mathcal{O}^{e}_{\varepsilon}}z_B$ for $e=(u,v)$. Additionally, the optimal objective value $Z^{\delta}_{E,\gamma}$ satisfies
\begin{align*}
    Z^{\delta}_{E,\gamma}=\delta\cdot \sum_{e\in E}w(e)\cdot x^*(e)+\sum_{v\in V} y_v +\sum_{B\in \mathcal{O}_{\varepsilon}}z_B\cdot \frac{\card{B}-1}{2}.
\end{align*}
\end{lemma2}
\begin{proof}
We consider the Lagrangian of the problem defined in \Cref{def:entropyregmatching}:
\begin{align*}   L(\vec{x},\vec{y},\vec{z},\vec{r})=g(\vec{x})+\sum_{v\in V}y_v\cdot(1-\sum_{e\ni v}x(e))+\sum_{B\in \mathcal{O}_{\varepsilon}}z_B\cdot \paren{\left\lfloor\frac{\card{B}}{2}\right\rfloor-\sum_{e\in G[B]}x(e)}+\sum_{e\in E} x(e)\cdot r_e.
\end{align*}
Suppose $(\vec{y}, \vec{z},\vec{r})$ is the optimal solution of the dual of the problem defined in \Cref{def:entropyregmatching}. Then by the KKT Stationarity Condition, we have, that the optimal primal solution is:
\begin{align*}
    x^*(e)=2^{\frac{1}{\delta}-1-\frac{s_e}{\delta\cdot w(e)}+\frac{r_e}{\delta\cdot w(e)}+\log\frac{\gamma}{w(e)}}
\end{align*}
Since $x^*(e)>0$ for all $e$, using complementary slackness, we have $r_e=0$ for all $e$. This proves the first part of the claim. Additionally, the \emph{entropy regularized matching problem} satisfies Slater's condition, and consequently has strong duality. Therefore, $Z_{E,\gamma}^{\delta}=L(\vec{x}^*,\vec{y},\vec{z},\vec{r})$. By substituting $(\vec{x}^*,\vec{y},\vec{z},\vec{r})$ into $L$, we have, 
\begin{align*}
    Z^{\delta}_{E,\gamma}=\delta\cdot\sum_{e\in E} w(e)\cdot x^*(e)+\sum_{v\in V}y_v+\sum_{B\in \mathcal{O}_{\varepsilon}}z_B\cdot\frac{\card{B}-1}{2}
\end{align*}
\end{proof}

Similarly, we have the following corollary, again adapted from \cite{CST23}.
\begin{corollary2}\label{cor:primalupperboundsdual}
    It holds that $Z^{\delta}_{E,\gamma}\geqslant \sum_{v\in V}y_v+\sum_{B\in \mathcal{O}_{\varepsilon}}z_B\cdot \frac{\card{B}-1}{2}$. 
\end{corollary2}

The next observation of \cite{CST23} states that the optimal dual solution $(\vec{y},\vec{z})$ corresponding to entropy regularized matching over $\mathcal{M}_{G,\varepsilon}$ is also a fractional cover for all edges. Given a fractional matching $\vec{x}$, we define $E_{\varepsilon}(x)=\set{e\in E\mid x(e)\geqslant \frac{\varepsilon}{3\cdot n}}$.

\begin{lemma2}\label{lem:dualprimal}
Let $\varepsilon>0$, let $w^*\leqslant \gamma\leqslant m\cdot w^*$, and suppose $\delta\leqslant \frac{\varepsilon}{8\cdot \log\paren{\frac{n^4\cdot W}{\varepsilon}}}$. Let $\vec{y},\vec{z}$ be the optimal dual solution to the entropy regularized matching problem over $\mathcal{M}_{G,\varepsilon}$. Then, for every edge $e\in E$, $s_e\geqslant (1-\varepsilon)\cdot w(e)$. Moreover, for every edge $e\in E_{\varepsilon}(x^*)$, $s_e\leqslant (1+\varepsilon)\cdot w(e)$. 
\end{lemma2}
\begin{proof}
    To prove the first part, we invoke \Cref{optimalprimal}, and the fact that $x^*(e)\leqslant 1$, we have,
    \begin{align*}
        \frac{1}{\delta}-1-\frac{s_e}{\delta\cdot w(e)}+\log\frac{\gamma}{w(e)}&\leqslant 0\\
        w(e)-\delta\cdot w(e)+\delta\cdot w(e)\cdot \log\frac{\gamma}{w(e)}&\leqslant s_e\\
        w(e)\cdot(1-\varepsilon)\leqslant s_e
    \end{align*}
The last inequality follows from the fact that $\gamma\geqslant w(e)$. For $e\in E_{\varepsilon}(x^*)$, we have:
\begin{align*}
    \log\frac{\varepsilon}{3n}&\leqslant \frac{1}{\delta}-1-\frac{s_e}{\delta\cdot w(e)}+\log\frac{\gamma}{w(e)}\\
    s_e&\leqslant w(e)\cdot(1-\delta)+\delta\cdot w(e)\log\frac{\gamma}{w(e)}-\delta\cdot w(e)\cdot \log\frac{\varepsilon}{n}\\
    &\leqslant (1+\varepsilon)\cdot w(e)
\end{align*}
This proves our claim.
\end{proof}
Similar to \cite{CST23} did for \textbf{modified entropy regularization problem}, we prove the following lemma for the unmodified version. 
\begin{lemma2}\label{lem:preservemostweight}
Let $\varepsilon>0$. Consider any fractional matching $\vec{x}\in \mathcal{M}_{G,\varepsilon}$, $w^*\leqslant \gamma\leqslant m\cdot w^*$, and $\delta\leqslant \frac{\varepsilon}{8\log\paren{\frac{n^4\cdot W}{\varepsilon}}}$, we have, 
\begin{align*}
    \sum_{e\in E_{\varepsilon}(x)}x(e)\cdot w(e)\geqslant \sum_{e\in E}x(e)\cdot w(e)-\varepsilon\cdot Z^{\delta}_{E,\gamma}.
\end{align*}
\end{lemma2}
\begin{proof}
For edges in $E\setminus E_{\varepsilon}(x)$, we have,
\begin{align*}
    (1-\varepsilon)\sum_{e\in E\setminus E_{\varepsilon}(x)}w(e)\cdot x(e)&\leqslant \sum_{e\in E\setminus E_{\varepsilon}(x)}s_e\cdot x(e)\\
    &\text{(From \Cref{lem:dualprimal})}\\
    &=\sum_{v\in V}y_v\cdot\sum_{e\in E_v\setminus E_{\eps}(x)}x(e)+\sum_{B\in \mathcal{O}_{\varepsilon}}z_B\cdot\sum_{e\in G[B]\setminus E_{\varepsilon}(x)} x(e)\\
    &\leqslant\varepsilon\cdot \sum_{v\in V}y_v+\varepsilon\cdot\sum_{B\in \mathcal{O}_{\varepsilon}}z_B\cdot \frac{\card{B}-1}{2}\\
    &\leqslant \varepsilon\cdot Z^{\delta}_{E,\gamma}\\
    &\text{(From \Cref{cor:primalupperboundsdual})}
\end{align*}
The second last inequality follows from the following line of reasoning:
\begin{align*}
\sum_{e\in E_v\setminus E_{\varepsilon}(x)}x(e)&\leqslant \sum_{e\in E_v\setminus E_{\varepsilon}(x)} \frac{\varepsilon}{n}\leqslant \varepsilon, \text{ and}\\
\sum_{e\in G[B]\setminus E_{\varepsilon}(x)}x(e)&\leqslant \frac{\varepsilon}{n}\cdot\card{B}\cdot \frac{\card{B}-1}{2}\leqslant \varepsilon\cdot \frac{\card{B}-1}{2}
\end{align*}
This shows our claim. 
\end{proof}
\begin{lemma2}\label{lem:weightpreserve}
Suppose $\varepsilon>0$, $w^*\leqslant \gamma\leqslant m\cdot \gamma$, and $\delta\leqslant \frac{\varepsilon}{\log\paren{\frac{n^4\cdot W}{\varepsilon}}}$, suppose $\vec{x}\in \mathcal{M}_{G,\varepsilon}$ with,
\begin{align*}
    \sum_{e\in E}w(e)\cdot \card{x(e)-x^*(e)}\leqslant \varepsilon\cdot Z^{\mu}_{E,\gamma},
\end{align*}
for any subset $E'\subset E$ such that,
\begin{align*}
    \sum_{e\in E'}w(e)\cdot x(e)\geqslant (1-\varepsilon)\cdot \sum_{e\in E}w(e)\cdot x(e),
\end{align*}
then suppose $\vec{f}$ is $\vec{x}$ restricted to $E_{\varepsilon}(x)$ and let $S$ be the output of \textsc{Sparsification}($\vec{f},\varepsilon,d$). Then, $S$ satisfies,
\begin{align*}
    \textsf{mwm}(S)\geqslant (1-\varepsilon)\cdot \sum_{e\in E'}w(e)\cdot x(e).
\end{align*}  
\end{lemma2}
\begin{proof}
    Let $S$ denote the output of $\textsc{Sparsification}(\vec{y},\varepsilon,d)$. Then, by \Cref{cor:sparsificationandweakdegreesparse}, we know that $\vec{x}^S$ is a $O(\log n)$-sparse, weak $\varepsilon$-degree sparsifier. Then, we want to show the following:
    \begin{align*}
        \sum_{e\in E'\cap E_{\varepsilon}(x)}w(e)\cdot x^{S}(e)\geqslant (1-\varepsilon)\cdot \sum_{e\in E'}w(e)\cdot x(e)
    \end{align*}
    Let $\vec{x}^*\in \mathcal{M}_{G,\varepsilon}$ be the optimal for the problem defined in \Cref{def:entropyregmatching}. Then, we have the following two observations from \Cref{lem:preservemostweight}:
    \begin{align*}
    \sum_{e\in E'\setminus E_{\varepsilon}(x)} w(e)\cdot x(e)&\leqslant \varepsilon\cdot Z^{\delta}_{E',\gamma}\leqslant 2\cdot\varepsilon\cdot Z^{\delta}_{E,\gamma}\\
    \sum_{e\in E'\setminus E_{\varepsilon}(x^*)}w(e)\cdot x^*(e)&\leqslant \varepsilon\cdot Z^{\delta}_{E',\gamma}\leqslant 2\cdot\varepsilon\cdot Z^{\delta}_{E,\gamma}
    \end{align*}
Using the premise of the lemma, we have, 
\begin{align*}
    \sum_{e\in E'\setminus E_{\varepsilon}(x^*)} w(e)\cdot x(e)&\leqslant \sum_{e\in E'\setminus E_{\varepsilon}(x^*)}w(e)\cdot x^*(e)+\sum_{e\in E'\setminus E_{\varepsilon}(x^*)} w(e)\cdot \card{x(e)-x^*(e)}\\
    &\leqslant 3\cdot\varepsilon\cdot Z^{\delta}_{E',\gamma}
\end{align*}
Consequently, we have,
\begin{align}\label{eqn:upperbdonlowweight}
\sum_{e\in E'\setminus E_{\varepsilon}(\vec{x}^*)\cap E_{\varepsilon}(\vec{x})} w(e)\cdot x(e)\leqslant 4\cdot\varepsilon\cdot Z^{\delta}_{E',\gamma}.
\end{align}
We now focus on proving the main statement mentioned above. We define $E'_{\varepsilon}(\vec{x}^*)\coloneqq E_{\varepsilon}(\vec{x}^*)\cap E'$, and similarly we define $E'_{\varepsilon}(\vec{x})$. Additionally, define $E'_{\varepsilon}(v)=E_{v}\cap E'_{\varepsilon}(\vec{x})\cap E'_{\varepsilon}(\vec{x}^*)$, and $G'_{\varepsilon}[B]=G[B]\cap E'_{\varepsilon}(\vec{x})\cap E'_{\varepsilon}(\vec{x}^*)$.
\begin{align*}
    \sum_{e\in E'}w(e)\cdot x^S(e)&\geqslant \sum_{e\in E'_{\varepsilon}(\vec{x})\cap E'_{\varepsilon}(\vec{x}^*)} w(e)\cdot x^S(e)\\
    &\geqslant (1-\varepsilon)\cdot \sum_{e\in E'_{\varepsilon}(\vec{x})\cap E'_{\varepsilon}(\vec{x}^*)}s_e\cdot x^S(e)\\
    &\text{(From \Cref{lem:dualprimal})}\\
    &\geqslant (1-\varepsilon)\cdot\sum_{v\in V}y_v\cdot\sum_{e\in E'_{\varepsilon}(v)} x^S(e)+\sum_{B\in \mathcal{O}_{\varepsilon}}z_B\cdot\sum_{e\in G'_{\varepsilon}[B]}x^S(e)\\
    &\geqslant (1-\varepsilon)\cdot\paren{\sum_{v\in V}y_v\sum_{e\in E'_{\varepsilon}(v)} x(e)+\sum_{B\in \mathcal{O}_{\varepsilon}}z_B\sum_{e\in G'_{\varepsilon}[B]}x(e)}-\varepsilon\cdot \paren{\sum_{v\in V}y_v+\sum_{B\in \mathcal{O}_{\varepsilon}}z_B\cdot\frac{\card{B}-1}{2}}\\
    &\text{(From \Cref{cor:sparsificationandweakdegreesparse} and \Cref{weakdegsparsifier})}\\
    &\geqslant (1-\varepsilon)\cdot \sum_{e\in E'_{\eps}(\vec{x})\cap E'_{\varepsilon}(\vec{x}^*)} w(e)\cdot x(e)-\varepsilon\cdot Z^{\delta}_{E,\gamma}\\
    &\text{(From \Cref{cor:primalupperboundsdual})}\\
    &\geqslant (1-\varepsilon)\cdot \sum_{e\in E'}w(e)\cdot x(e)-3\cdot \varepsilon\cdot Z^{\delta}_{E,\gamma}\\
    &\text{(From \Cref{eqn:upperbdonlowweight})}\\
    &\geqslant (1-3\varepsilon)\cdot \sum_{e\in E'}w(e)\cdot x(e)
\end{align*}
\end{proof}

\begin{fact2}\label{prop:entropyregmatching}
Suppose $\vec{z}\in \mathcal{M}_{G,\varepsilon}$, and $\vec{x}^*$ be the optimal solution to the \emph{entropy regularized matching problem} Then, 
\begin{align*}
g(\vec{z})\leqslant g(\vec{x}^*)-\frac{\delta}{\textsf{mwm}(G)}\cdot \paren{\sum_{e\in E}w(e)\cdot\card{z(e)-x^*(e)}}^2.
\end{align*}
\end{fact2}
\begin{lemma2}\label{lem:closeopt}
Suppose $\vec{y}\in \mathcal{M}_{G,\varepsilon}$ is a $(1-\varepsilon)$-approximate solution to the \textbf{entropy regularization problem}, and $\vec{x}^*\in \mathcal{M}_{G,\varepsilon}$ is the optimal solution to the \textbf{entropy regularization} problem. Then,
\begin{align*}
     \sum_{e\in E}w(e)\cdot \card{y(e)-x^*(e)}\leqslant \varepsilon\cdot Z^{\delta}_{E,\gamma}
\end{align*}
\end{lemma2}
\begin{proof}
    Observe that we have, $g(\vec{y})\geqslant (1-\varepsilon)\cdot g(\vec{x}^*)$. Consequently, we have, $g(\vec{x}^*)-g(\vec{y})\leqslant \varepsilon\cdot Z^{\delta}_{E,\gamma}$. Then, combining the fact that $Z^{\delta}_{E,\gamma}\geqslant \textsf{mwm}(G)$ and \Cref{prop:entropyregmatching}, we have, 
    \begin{align*}
        \varepsilon\cdot Z^{\delta}_{E,\gamma}&\geqslant \frac{\delta}{\textsf{mwm}(G)}\cdot \paren{\sum_{e\in E}w(e)\cdot \card{z(e)-x^*(e)}}^2\\
        \varepsilon\cdot Z^{\delta}_{E,\gamma}&\geqslant \sum_{e\in E}w(e)\cdot\card{z(e)-x^*(e)}
    \end{align*}
\end{proof}

Thus, from \Cref{lem:closeopt} and \Cref{lem:preservemostweight}, we have concluded the proof that the entropy regularized fractional matching when fed into \textsc{Sparsification}() outputs a sparsifier $S$ that preserves most of its weight.

\subsection{From Approximate Entropy Regularized Matching to Decremental Matching}

In this section, show how to get decremental matching algorithms from \textbf{approximate entropy regularized matching}. The proof is almost identical to that given by \cite{CST23}. We just state it here for completeness. We start with the following definition, used by \cite{CST23}:

\begin{definition2}\label{def:modifiedentropy}
    The \textbf{modified entropy regularized matching} problem is the problem of finding a fractional matching $\vec{x}\in \mathcal{M}_G$, maximizing the following objective function:
    \begin{align*}
        g(\vec{x})=\sum_{e\in E}x(e)\cdot w(e)+\delta\cdot\sum_{e\in E}w(e)\cdot x(e)\cdot\log\frac{\gamma}{w(e)\cdot x(e)}
    \end{align*}
Let $Y^{\delta}_{E,\gamma}$ denote the optimal solution to this problem.
\end{definition2}
\begin{observation2} \label{obs:relbwmodifiedandunmodified}
    We have, $Y^{\delta}_{E,\gamma}\geqslant Z^{\delta}_{E,\gamma}\cdot(1-\varepsilon)$.
\end{observation2}
\begin{proof}
    Suppose $\vec{x}^*\in\mathcal{M}_{G,\varepsilon}$ realizes $Z^{\delta}_{E,\gamma}$. From \Cref{obs:smallblossomsuff}, we have, $\vec{z}=\frac{\vec{x}^*}{1+\varepsilon}\in \mathcal{M}_{G}$. Thus, we have,
    \begin{align*}
        Y^{\delta}_{E,\gamma}\geqslant g(\vec{z})\geqslant (1-\varepsilon)\cdot g(\vec{x}^*)\geqslant Z^{\delta}_{E,\gamma}\cdot (1-\varepsilon).
    \end{align*}
This proves the observation.
\end{proof}

The next lemma shows that we can compute an approximate solution to the modified entropy maximization problem in $O(m\cdot \poly(\log n, \nicefrac{1}{\varepsilon}))$ time. 

\begin{lemma2}\cite{CST23}\label{lem:guaranteesEntropyReg}
For any $\varepsilon=\Omega(\nicefrac{1}{\sqrt{n}})$, there is a randomized algorithm \textsc{EntRegMatching}() that give a $(1-\varepsilon)$-approximation to the \textbf{modified entropy regularized matching problem} in $O(m\cdot\varepsilon^{-6}+n\cdot \varepsilon^{-13})$ time. 
\end{lemma2}

Using the above lemma, we will show the following theorem:

\densegraphs*

We also need the following lemma, which states that the solution to the \emph{modified entropy regularized matching} is a $(1-\varepsilon)$-approximation to $\textsf{mwm}(G)$. 

\begin{lemma2}[\cite{CST23}]\label{lem:propapproxEntropyRegMatching}
For $\mu\leqslant \frac{\varepsilon}{\log n\cdot W}$ and $\textsf{mwm}(G)\leqslant \gamma \leqslant n\cdot W\cdot \textsf{mwm}(G)$, let $\vec{x}^*$ be the optimal solution to the \emph{modified entropy regularized matching} problem, then we have,
\begin{align*}
    \sum_{e\in E} w(e)\cdot x^*(e)\geqslant (1-\varepsilon)\cdot \textsf{mwm}(G).
\end{align*}
Similarly, suppose $\vec{x}$ is a $(1-\varepsilon)$-approximate solution to the \emph{modified entropy regularized matching} problem, then we have,
\begin{align*}
    \sum_{e\in E} w(e)\cdot x(e)+\varepsilon\cdot \textsf{mwm}(G)\geqslant (1-\varepsilon)\cdot g(\vec{x}^*)\geqslant (1-\varepsilon)\cdot \sum_{e\in E} w(e)\cdot x^*(e)\geqslant (1-\varepsilon)^2\cdot \textsf{mwm}(G)
\end{align*}
\end{lemma2}

In order to prove \Cref{thm:main}, we state the main algorithm, and then show its correctness and runtime.

\begin{algorithm}
\caption{\textsc{Decremental Matching}}
\algorithmicrequire{ $G$, weights $w$, and a precision parameter $\varepsilon>0$}
\begin{algorithmic}[1]
\Procedure{Initialize}{$G,E$}
\State Initialize $G\leftarrow G_0$ \Comment{$G$ refers to the current graph, and $G_0$ is the initial graph.}
\State Initialize $E\leftarrow E_0$ \Comment{$E$ refers to the current edge set, and $E_0=E(G_0)$.}
\State \textsc{CounterM}$\leftarrow 0$\Comment{Counts the total weight deleted from $M$}
\State \textsc{CounterX}$\leftarrow 0$\Comment{Counts the total weight deleted from $\vec{x}$}
\EndProcedure
\Procedure{\textsc{Rebuild}}{}
\State $\vec{y}\leftarrow $\textsc{EntRegMatch}($G,\varepsilon$)
\State Let $\vec{x}$ be $\vec{y}$ restricted to $E_{\varepsilon}(\vec{y})$.
\State $\mu^*\leftarrow \sum_{e\in E}w(e)\cdot x(e)$ \Comment{$\mu^*\geqslant (1-\varepsilon)\cdot \textsf{mwm}(G)$}
\EndProcedure
\Procedure{\textsc{Round}}{}
\State $S\leftarrow \textsc{Sparsification}(\vec{x},\varepsilon)$
\State $M\leftarrow \textsc{Static-Match}(S,\varepsilon)$
\EndProcedure
\Procedure{Deletion}{$e$}
\State $E\leftarrow E\setminus e$
\If{$e\in \text{supp}(\vec{x})$}
\State Delete $e$ from \text{supp}($\vec{x}$)
\State $\textsc{CounterX}\leftarrow \textsc{CounterX}+w(e)\cdot x(e)$
\EndIf
\If{$\textsc{CounterX}\geqslant \varepsilon\cdot \mu^*$}
\State \textsc{Rebuild}()
\EndIf
\If{$e\in M$}
\State $M\leftarrow M\setminus \set{e}$
\State $\textsc{CounterM}\leftarrow \textsc{CounterM}+1$
\EndIf
\If{$\textsc{CounterM}\geqslant \varepsilon\cdot \mu^*$}
\State $\textsc{CounterM}\leftarrow 0$
\State \textsc{Round}()\label{line:roundwhenMdrops}
\EndIf
\EndProcedure
\end{algorithmic}
\label{alg:decremental}
\end{algorithm}

\begin{lemma2}
\Cref{alg:decremental} maintains an integral matching $M$ at all times such that $w(M)\geqslant (1-\varepsilon)\cdot \textsf{mwm}(G)$.
\end{lemma2}
\begin{proof}
    Note that from \Cref{lem:guaranteesEntropyReg}, we can conclude that the fractional matching $\vec{y}$ output by \textsc{EntRegMatch}($G,\varepsilon$) has the property that:
    \begin{align*}
        g(\vec{y})\geqslant (1-\varepsilon)\cdot g(\vec{y}^*),
    \end{align*}
where $\vec{y}^*\in \mathcal{M}_G$ is the optimal solution to the \textbf{modified entropy regularization} problem. This satisfies the premise of \Cref{lem:propapproxEntropyRegMatching}. Consequently, from that we can deduce:
\begin{align*}
\sum_{e\in E}w(e)\cdot y(e)\geqslant (1-\varepsilon)^2\cdot \textsf{mwm}(G).
\end{align*}
Note that $\vec{x}$ is $\vec{y}$ restricted to $E_{\varepsilon}(\vec{y})$. Therefore, from \Cref{lem:preservemostweight} and \Cref{obs:entropyregmwm}, we can then conclude the following:
\begin{align*}
    \sum_{e\in E}w(e)\cdot x(e)\geqslant \sum_{e\in E}w(e)\cdot y(e)-\varepsilon\cdot Z^{\delta}_{E,\gamma}\geqslant (1-3\varepsilon)\cdot \textsf{mwm}(G)
\end{align*}
Additionally, note that the algorithm processes deletions, and rebuilds when $\sum_{e\in E}w(e)\cdot x(e)$ has dropped by a $(1-\varepsilon)$ factor. Thus, each time the algorithm is maintaining a fractional matching $\vec{x}$ such that $\sum_{e\in E}w(e)\cdot x(e)\geqslant (1-\varepsilon)^3\cdot \textsf{mwm}(G)$. Next, the algorithm runs \textsc{Sparsification}($\vec{x},\varepsilon$) to compute a sparsifier $S$. Observe $\vec{y}$ satisfies the premise of \Cref{lem:closeopt}. Consequently, we have, \Cref{obs:relbwmodifiedandunmodified},
\begin{align*}
    \sum_{e\in E}w(e)\cdot \card{y(e)-x^*(e)}\leqslant \varepsilon\cdot Z^{\delta}_{E,\gamma}
\end{align*}
Note that we feed $\vec{x}$, which is $\vec{y}$ restricted to $E_{\varepsilon}(\vec{y})$ into \textsc{Sparsification}(). Thus, from \Cref{lem:weightpreserve}, we have,
\begin{align*}
    \textsf{mwm}(S)\geqslant (1-\varepsilon)\cdot \sum_{e\in E}w(e)\cdot y(e)
\end{align*}
Therefore, from \Cref{lem:duanpettie} we can conclude that the matching $M$ output by $\textsc{Static-Match}(S,\varepsilon)$ has $w(M)\geqslant\geqslant (1-\varepsilon)\cdot \textsf{mwm}(S)\geqslant (1-\varepsilon)^2\cdot\sum_{e\in E}w(e)\cdot y(e)\geqslant (1-\varepsilon)^3\cdot \textsf{mwm}(G)$. Here, \Cref{lem:propapproxEntropyRegMatching} gives us the last inequality. This shows the lemma.
\end{proof}

Now, we show the runtime, and for that we need the following lemma.

\begin{lemma2}\label{lem:dropinentropy}(\cite{CST23})
If \textsc{EntRegMatch}() returns a $(1-\varepsilon)$ approximate solution to $g(\vec{x})$, then the \textsc{Rebuild}() will be called at most $O(\poly(\log n,\nicefrac{1}{\varepsilon}))$ times before \textsf{mwm}($G$) drops from at least $\mu^*$ to atmost $\mu^*\cdot (1-\varepsilon)$. 
\end{lemma2}

\begin{lemma2}
\Cref{alg:decremental} has an amortized updated time of $\Tilde{O}\paren{\poly(\nicefrac{1}{\varepsilon})+\frac{n^2}{m}\cdot \poly(\nicefrac{1}{\varepsilon})}$.
\end{lemma2}
\begin{proof}
First, observe the runtime of each of the procedures and each time they are invoked:
\begin{enumerate}
\item For the procedure \textsc{Rebuild}(), we can conclude from \Cref{lem:guaranteesEntropyReg} that each time it is run it takes time $\Tilde{O}(m\cdot \poly(\log n,\nicefrac{1}{\varepsilon}))$. This procedure is called each time \textsc{CounterX} increases to $\varepsilon\cdot \mu^*$ or $\sum_{e\in E^*}w(e)\cdot x(e)$ has dropped by $(1-\varepsilon)$ amount. From \Cref{lem:dropinentropy}, we can conclude \textsc{Rebuild}($\vec{x},\varepsilon$) is called at most $O(\poly(\log n,\nicefrac{1}{\varepsilon}))$ times. Thus, the total runtime of this procedure over $m$ deletions is at most $O(m\cdot\poly(\log n,\nicefrac{1}{\varepsilon}))$. That is, till the matching weight drops from $\mu^*$ to $1$. 

\item Next we consider the procedure \textsc{Round}(). Note that from \Cref{obs:sizeofsparsifier} we can conclude that $\card{E(S)}= O(n\cdot\poly(\log n,\nicefrac{1}{\varepsilon}))$. Moreover, the time taken to maintain it is $O(\poly(\log n,\nicefrac{1}{\varepsilon}))$. Thus, from \Cref{lem:duanpettie} we can conclude that \textsc{Static-Match}($S,\varepsilon$) and consequently, the procedure $\textsc{Round}()$ take time $O(n\cdot \poly(\log n,\nicefrac{1}{\varepsilon}))$ each time they are invoked. This is invoked each time \textsc{CounterM} drops by $\varepsilon\cdot \mu^*$ amount. Since $x(e)\geqslant \nicefrac{\varepsilon}{n}$ for all $e\in E$, this also translates into the adversary deleting a weight of at least $\nicefrac{\varepsilon^2\cdot \mu^*}{n}$ from the fractional matching $\vec{x}$. Consequently, till before the next time \textsc{Rebuild}() is invoked, \textsc{Round}() is invoked $O(\nicefrac{n}{\varepsilon^2})$ times. Since the total number of times \textsc{Rebuild}() is called is at most $O(\poly(\log n,\nicefrac{1}{\varepsilon}))$, this implies that the total number of times $\textsc{Round}()$ is called is at most $O(n\cdot\poly(\log n,\nicefrac{1}{\varepsilon}))$, and therefore, the total runtime contributed by this procedure is $O(n^2\cdot \poly(\log n,\nicefrac{1}{\varepsilon}))$.  
\item The other steps in the \textsc{Deletion}() procedure take $O(1)$ time per edge to implement. So the total time over all updates is at most $O(m)$.
\end{enumerate}
This concludes the proof.
\end{proof}

\section{Acknowledgements}
Thank you to David Wajc for numerous helpful email exchanges. Thank you to Rishabh Dudeja, Madhusudhan Raman and Aditya Potukuchi for encouragement. 
\bibliographystyle{alpha}
\bibliography{references}
\end{document}